%% file: sampler.tex
\documentclass[manuscript, letterpaper]{aastex6}

\include{gitstuff}
\include{aastexmods}

\definecolor{cbblue}{HTML}{3182bd}
\usepackage{microtype}  
\usepackage{amsmath}
\hypersetup{backref,breaklinks,colorlinks,urlcolor=cbblue,linkcolor=cbblue,citecolor=black}

\newcommand{\project}[1]{\textsl{#1}}
\newcommand{\acronym}[1]{{\small{#1}}}
\newcommand{\apogee}{\project{\acronym{APOGEE}}}
\newcommand{\sdssiii}{\project{\acronym{SDSS-III}}}
\newcommand{\samplername}{\project{The~Joker}}
\newcommand{\emcee}{\project{emcee}}

\newcommand{\documentname}{\textsl{Article}}
\newcommand{\sectionname}{Section}
\newcommand{\figname}{Figure}
\newcommand{\eqname}{Equation}

\newcommand{\kms}{\mathrm{km}~\mathrm{s}^{-1}}
\newcommand{\mps}{\mathrm{m}~\mathrm{s}^{-1}}
\newcommand{\bs}[1]{\boldsymbol{#1}}

\definecolor{mahogany}{RGB}{165,15,21}
\newcommand{\resp}[1]{#1}

\newcounter{expcounter}
\stepcounter{expcounter}

\begin{document}\sloppy\sloppypar\raggedbottom\frenchspacing 

\title{\samplername: A custom Monte Carlo sampler \\
  for binary-star and exoplanet radial velocity data}
\author{Adrian~M.~Price-Whelan\altaffilmark{\pu,\adrn},
        David~W.~Hogg\altaffilmark{\ccpp,\cds,\mpia,\flatiron},
        Daniel~Foreman-Mackey\altaffilmark{\uw,\sagan},
        Hans-Walter~Rix\altaffilmark{\mpia}
}

\newcommand{\pu}{1}
\newcommand{\adrn}{2}
\newcommand{\ccpp}{3}
\newcommand{\cds}{4}
\newcommand{\mpia}{5}
\newcommand{\flatiron}{6}
\newcommand{\uw}{7}
\newcommand{\sagan}{8}

\altaffiltext{\pu}{Department of Astrophysical Sciences,
                   Princeton University, Princeton, NJ 08544, USA}
\altaffiltext{\adrn}{To whom correspondence should be addressed:
                     \texttt{adrn@astro.princeton.edu}}
\altaffiltext{\ccpp}{Center for Cosmology and Particle Physics,
                     Department of Physics,
                     New York University, 726 Broadway,
                     New York, NY 10003, USA}
\altaffiltext{\cds}{Center for Data Science,
                     New York University, 60 Fifth Avenue,
                     New York, NY 10011, USA}
\altaffiltext{\mpia}{Max-Planck-Institut f\"ur Astronomie,
                     K\"onigstuhl 17, D-69117 Heidelberg, Germany}
\altaffiltext{\flatiron}{Flatiron Institute,
                         Simons Foundation,
                         162 Fifth Avenue,
                         New York, NY 10010, USA}
\altaffiltext{\uw}{Astronomy Department, University of Washington,
                   Seattle, WA 98195, USA}
\altaffiltext{\sagan}{Sagan Fellow}

\begin{abstract}
Given sparse or low-quality radial-velocity measurements of a star, there are
often many qualitatively different stellar or exoplanet companion orbit models
that are consistent with the data.
The consequent multimodality of the likelihood function leads to extremely
challenging search, optimization, and MCMC posterior sampling over the orbital
parameters.
\resp{Here we create a custom Monte Carlo sampler for sparse or noisy
radial-velocity measurements of two-body systems that can produce posterior
samples for orbital parameters even when the likelihood function is poorly
behaved.}
The six standard orbital parameters for a binary system can be split into four
non-linear parameters (period, eccentricity, argument of pericenter, phase)
and two linear parameters (velocity amplitude, barycenter velocity).
We capitalize on this by building a sampling method in which we densely sample
the prior pdf in the non-linear parameters and perform rejection sampling using
a likelihood function marginalized over the linear parameters.
With sparse or uninformative data, the sampling obtained by this rejection
sampling is generally multimodal and dense.
With informative data, the sampling becomes effectively unimodal but too
sparse: in these cases we follow the rejection sampling with standard MCMC.
The method produces correct samplings in orbital parameters for data that
include as few as three epochs.
\samplername\ can therefore be used to produce proper samplings of multimodal
pdfs, which are still informative and can be used in hierarchical
(population) modeling.
We give some examples that show how the posterior pdf depends sensitively on the
number and time coverage of the observations and their uncertainties.
\end{abstract}


\section{Introduction} \label{sec:intro}

Precise radial-velocity measurements of stars have transformed
astrophysics in the last decades.
They have permitted the discovery and confirmation of companions (planetary,
stellar, and otherwise) orbiting thousands of stars.
Radial velocity surveys hold the promise of delivering the full population
statistics for binary and trinary star systems (for example,
\citealt{Raghavan:2010, Tokovinin:2014, Troup:2016}).

With large stellar spectroscopic surveys operating or under
construction, we expect to have good quality spectra for millions
of stars in the next few years.
Most of these surveys have at least some targets---and many have many
targets---that are observed multiple times (e.g., \citealt{Majewski:2015}).
Whether as a primary or secondary goal of their observing
strategies, these surveys can generate discoveries of planetary,
substellar, and stellar companions.
These discoveries, in turn, will feed population inferences, follow-up
programs, and projects to refine precise stellar models.

However, when radial-velocity observations are not designed with
unambiguous detection and discovery in mind, there are usually
multiple possible star-companion (orbit) models that are consistent with any
modest number of radial-velocity measurements that show stellar
acceleration.
That is, a small number of radial velocity measurements---even when the
uncertainties are small---will lead to posterior beliefs about companion
orbits and masses that put substantial plausibility onto multiple
qualitatively different solutions.
This is then reflected in a likelihood function that is highly multi-modal in
the relevant parameter spaces (e.g., Keplerian orbital parameters).
While multi-modal orbit likelihoods may be frustrating when studying individual
systems of particular interest, extensive sets of such likelihoods can
be powerful constraints for hierarchical modeling, inferring, e.g., the
characteristics of the binary star, or exoplanet population.

While the problem has of course been recognized for a long time, there are
currently no methods that explore these highly multimodal
\resp{distributions} with good guarantees of \resp{converged samplings} (but see
\citealt{Gregory:2005,Brewer:2015}).
\resp{Converged, independent samplings are essential to the jobs of delivering
correct posterior samplings and reliable probabilistic statements about
detection and characterization.
With general Markov Chain Monte Carlo (MCMC) methods, the returned samplings are
inherently correlated and must therefore be run longer than the autocorrelation
time.
However, computing the autocorrelation time (a two-point statistic of a chain)
is not simple to compute and can be misleading, especially when the target
distribution is highly multimodal with widely separated modes.}

Here we present a path to address this problem.
Our approach is to build a custom posterior sampling method that capitalizes on
the structure of the binary-star (or star-exoplanet) kinematics to generate
robust samplings of multiple solutions at manageable computational cost.
In what follows, we use the term ``binary'' for any system with an observed
source (the primary, e.g., a star), whose radial velocity variations are
explained through gravitational two-body interactions with another object (the
companion, e.g., star, exoplanet, stellar remnant); i.e. we restrict our
analysis to single-line spectroscopic binary systems.

The structure of this \documentname\ is as follows:
We lay out our assumptions and implementation approach, and demonstrate that we
have a method that is reliable and essentially always correct under
those assumptions.
We perform experiments with the method to understand its properties and
limitations.
We then discuss the astrophysical applicability and potential of the
method.
We finish by describing the changes we would have to make if we weakened our assumptions, or
if we don't weaken our assumptions but they indeed prove to be far from correct.

\section{Assumptions and method} \label{sec:method}

In order to set up a well-posed problem and build a path to a definite solution,
we make a set of sensible, but non-trivial assumptions about the stellar systems we will use
and observations thereof.
Ultimately, we assume that the radial velocity curve of a single-line
spectroscopic binary system can be specified by six parameters
(\citealt{Kepler:1609}).
Here we adopt a parameterization for this problem that is similar to
\citealt{Murray:2010}:
$P$, $e$, $\phi_0$, $\omega$, $K$, and $v_0$, which are the period,
eccentricity, pericenter phase and argument, velocity semi-amplitude, and the
barycenter velocity.
As is well-known, this does not fully specify the binary system itself, because
of the inclination ($\sin{i}$) and mass function degeneracies.
To proceed, we assume the following:
\begin{itemize}\itemsep0ex  
\item We have measurements of the radial velocity of a
  star, and that the time dependence of the expectation of that radial
  velocity is well described by the gravitational orbit of a pair of
  point masses (the two-body problem). We assume that the uncertainties
  in the observation epochs are negligible, and specified in an inertial frame (for
  example, Solar-System barycentric Julian date).
  We only consider the case of a single-line spectroscopic binary system,
  where we do not have measurements of the (fainter) companion's projected orbit.
\item  Each star has exactly one companion, and that the radial-velocity
  measurements are not contaminated by nor affected by any other bodies. Our
  analysis allows the effective mass of the exactly-one companion to go to zero
  \resp{with finite probability}; this encompasses the case of no companion.
\item The noise contributions to individual radial-velocity measurements are
  well described as \resp{samples} from zero-mean normal (Gaussian)
  distributions with correctly known variances convolved with a zero-mean normal
  distribution with an additional ``jitter'' variance, $s^2$. We assume that
  there are no outliers beyond this flexible noise model.
\item We can put particular, fairly sensible but not highly restrictive,
  prior pdfs on all the orbital parameters, as we
  describe below.
\end{itemize}
Each of these assumptions could be challenged: in particular we expect some stars
to have additional companions, and we expect there to be outliers and
unaccounted sources of noise.
We will return to these assumptions, and the consequences of relaxing them, in
\sectionname~\ref{sec:discussion}.

The radial velocity $v$ at time $t$ is then given by (see also \eqname~63 in
\citealt{Murray:2010})
\begin{equation}
  v(t;\bs{\theta}) = v_0 + K\,[\cos(\omega + f) + e\,\cos\omega]
\end{equation}
where the $\bs{\theta}$ represents the free parameters, $f$ is the true anomaly
given by
\begin{equation}
  \cos f = \frac{\cos E - e}{1 - e\, \cos E}
\end{equation}
and the eccentric anomaly, $E$, must be solved for with the mean
anomaly, $M$,
\begin{align}
  M &= \frac{2\pi\, t}{P} - \phi_0\\
  M &= E - e\,\sin E \quad .
\end{align}
Of these parameters, four ($P$, $e$, $\omega$, $\phi_0$) are non-linearly
related to the radial-velocity expectation, and two ($K$, $v_0$) are
linearly related.
We additionally allow that the radial velocity curve of any star has an overall
jitter, $s^2$, to vary to partially account for
imperfect knowledge of the radial velocity uncertainties and any intrinsic
radial-velocity scatter; the jitter must also be treated as a non-linear
parameter.

With such a parameterization, the problem is then to construct the posterior pdf
for these seven parameters, accounting for the fact that this pdf
may have very complex, multi-modal structure.
Fundamentally, the method we describe and demonstrate here is to perform
rejection sampling in the non-linear parameters, but with analytic
marginalization over the two linear parameters.
The method capitalizes on the unique problem structure:
\begin{itemize}\itemsep0ex
\item There are both linear and non-linear parameters, and we can
  treat them differently; in particular, it is possible to
  analytically marginalize out the linear parameters (provided that the
  noise model is well-behaved and the prior pdf is conjugate).
\item There is a finite, time-sampling-imposed minimum size or
  resolution---in the period---of any features in the
  likelihood function. That is, there cannot be arbitrarily narrow
  modes in the multimodal posterior pdf.
\end{itemize}

\resp{The method described above is a specific case of} rejection-sampling
(\citealt{VonNeumann:1951}) in which we densely sample
(\resp{generate many samples with typical spacing smaller than the time-sampling
imposed resolution}) from the prior probability density function (prior pdf) and
use the likelihood evaluated at these samples as the rejection scalar.
In detail, the rejection step works as follows:
\begin{enumerate}\itemsep0ex
\item For each sample $j$ in the prior pdf sampling of the four non-linear
  parameters, there is a (linear, not logarithmic) \resp{marginal} likelihood
  value $L_j$ (a probability for the data given the \resp{non-linear}
  parameters).
\item There is a maximum value $L_{\rm max}$ that is the largest value of
  $L_j$ found across all of the samples in the prior sampling.
\item For each sample $j$, choose a random number $r_j$ between 0 and
  $L_{\rm max}$
\item Reject the sample $j$ if $L_j < r_j$.
\item The number of samples that survive the rejection is (hereafter) $M$.
\end{enumerate}
Note that this algorithm is guaranteed to produce at least one
surviving sample; of course if only one sample survives (or any very
small number), the sampling is not guaranteed to be fair.
That said, if the original sampling of the prior pdf is dense enough
that many survive the rejection step, the surviving samples do, by construction,
constitute a fair, uncorrelated sampling from the posterior pdf.

Our prior pdf in the non-linear parameters is very straightforward:
\begin{align}
    p(\ln P) &= \mathcal{U}(\ln P_{\rm min}, \ln P_{\rm max})\\
    p(e) &= {\rm Beta}(a, b) = \frac{\Gamma(a+b)}{\Gamma(a) \, \Gamma(b)} \, e^{a-1} \, [1 - e]^{b-1}\\
    p(\omega) &= \mathcal{U}(0, 2\pi) ~ [{\rm rad}]\\
    p(\phi_0) &= \mathcal{U}(0, 2\pi) ~ [{\rm rad}]\\
    p(\resp{\ln s}) &= \mathcal{N}(\mu_s, \sigma^2_s)
\end{align}
where $\mathcal{U}(x_1, x_2)$ is the uniform distribution over the domain $(x_1,
x_2)$, $\mathcal{N}(\mu, \sigma^2)$ is the normal distribution with mean $\mu$
and variance $\sigma^2$, and the prior pdf over eccentricity is the Beta
distribution with $a=0.867$, $b=3.03$ \citep{Kipping:2013}.
\resp{To simplify the marginalization integrals below, we assume that the priors
over the linear parameters ($K$, $v_0$) are very broad and Gaussian---or at
least very flat over the range of relevance---and that they do not depend on the
nonlinear parameters in any way (which is a substantial restriction; see
\sectionname~\ref{sec:discussion}).
Of course, we may actually have stronger prior beliefs about the systemic
velocity of the binary, $v_0$ (i.e. we may want to impose that it belong to the
Galactic disk, or use a mixture model for the different kinematic components of
the Galaxy).
Using a more informative prior would require only a minimal change to the
method.}

In \sectionname~\ref{sec:experiments} or in the subsections specific to the
different experiments we specify the values for hyperparameters $P_{\rm min}$,
$P_{\rm max}$, $\mu_s$, and $\sigma^2_s$.
In practice, the choice of $\mu_s$ and $\sigma^2_s$ can be tuned appropriately
depending on knowledge about intrinsic variability of the source or suspicions
about the reported uncertainties.
We sample the prior pdf directly and explicitly with standard
\project{numpy.random} calls (\citealt{Van-der-Walt:2011}).
In practice, we are usually required to take around $J=2^{28}$ samples (indeed,
a quarter billion samples) from the prior-pdf to produce sufficient final
samplings; in each experiment below, we state  the total number of prior samples
generated and the number of surviving samples.

The unmarginalized likelihood function $L$ is
\begin{equation}
\ln L = -\frac{1}{2}\,\sum_{n=1}^N \left[\frac{[v_n - v(t_n;\bs{\theta})]^2}{\sigma_n^2 + s^2}
 +\ln \left(2\pi\,[\sigma_n^2 + s^2]\right) \right]
\end{equation}
where $n$ indexes the individual data points $v_n$, $v(t)$ is the radial
velocity prediction at time $t$ given the orbital parameters $\bs{\theta}$, the
data-point times are the $t_n$, and $\sigma_n^2$ is the Gaussian noise variance
for data point $n$.
Note that the form of this likelihood function is fully specified by the
assumptions, given above.

We rejection-sample, however, using a marginalized likelihood, where we
analytically marginalize out the linear parameters ($K$, $v_0$).
We construct an $N \times 2$ design matrix consisting of a column of unit-$[K]$
predictions (given the non-linear parameters) and a column of ones.
We perform standard linear least-square fitting with this design
matrix to obtain the best-fit values for the two linear parameters,
and the standard $2\times 2$ linear-fitting covariance matrix $C_j$ for their
uncertainties.
With these, we can construct---for each prior sample---the marginalized
likelihood $Q_j$
\begin{equation}
\ln Q_j = -\frac{1}{2}\,\sum_{n=1}^N \left[\frac{[v_n - v(t_n;\bs{\theta}_j)]^2}{\sigma_n^2 + s^2}
 +\ln \left(2\pi\,[\sigma_n^2 + s^2]\right) \right] -\frac{1}{2}\,\ln ||2\pi\,C_j||
\end{equation}
where the prediction $v(t_n;\bs{\theta}_j)$ is taken at the best-fit values of
the linear parameters given sample $j$ of the non-linear parameters, and the
log-determinant term ($\ln ||C_j||$) accounts for the volume in the
marginalization integral.
These $Q_j$ are used in the rejection sampling algorithm described above.

There are three possible outcomes of this rejection sampling, based on two
thresholds:
We set a minimum number of samples $M_{\rm min}=128$.
We also set a period resolution $\Delta = [4\,P^2] / [2\pi\,T]$, with
$P$ set to the median period across the surviving samples and $T$ set to
the epoch span of the data.
This $\Delta$ is an expansion of the period resolution expected from an
information-theory (sampling theorems) perspective:
for a periodic signal with frequency $\omega$ observed over a window with size
$T$, the smallest resolvable frequency differences will be $\Delta\omega \approx
T^{-1}$, corresponding to period differences of $\Delta P \approx
\frac{P^2}{2\pi\,T}$.
The three possible outcomes are:
\begin{itemize}\itemsep0ex
\item $M\geq M_{\rm min}$ samples survive the rejection.
  In this case, we are done.
\item $M<M_{\rm min}$ samples survive the rejection, and these samples have a
  root-variance (rms) in the period parameter $P$ that is smaller than $\Delta$
  (i.e. they give no indication of period ambiguity).
  In this case we assume that the posterior pdf is effectively unimodal, and we
  use the surviving samples (or sample) to initialize a MCMC sampling using the
  \emcee\ package (\citealt{Foreman-Mackey:2013}).
\item $M<M_{\rm min}$ samples survive the rejection, and these samples
  span a period range larger than $\Delta$.
  In this case, we iterate the \resp{rejection-sampling} procedure: We generate
  new prior-pdf samplings and rejection sample until the number of surviving
  samples is larger than $M_{\rm min}$.
  This is expensive.
\end{itemize}
When we trigger the \resp{initialization} and operation of \emcee, we do the
following:
\begin{enumerate}\itemsep0ex
\item Randomly generate $M_{\rm min}$ sets of parameters $\bs{\theta}_m$
  (linear and nonlinear parameters) in a small, Gaussian ball around
  the highest-$Q_j$ sample from the \resp{rejection sampling}.
\item Run \emcee\ on this ensemble (with $M_{\rm min}$ walkers) for $2^{16}$
  steps \resp{(this number is arbitrary and can be tuned if the walkers converge
  much faster or slower)}.
\item Take the final state of the $M_{\rm min}$-element ensemble as an
  independent sampling of the posterior pdf.
\end{enumerate}
This procedure ensures that no matter what path we take, we end up with
at least $M_{\rm min}$ samples from the posterior for any input data.

Within the initial assumptions, this procedure almost inevitably results in a
correct sampling of the parameter pdf. This is ensured by the density of the
prior sampling in the nonlinear parameters, and also borne out in the numerical experiments
in the following \sectionname.

\section{Experiments and results} \label{sec:experiments}

In what follows, we use (1) simulated data with known properties and (2) actual
spectroscopic data from Data Release 13 (DR13;
\citealt{SDSS-Collaboration:2016}) of the Apache Point Observatory Galactic
Evolution Experiment (\apogee; \citealt{Majewski:2015}) in a series of
experiments that demonstrate the reliability and utility of \samplername.
\apogee\ is one of the four sub-surveys of the Sloan Digital Sky Survey-III
(SDSS-III; \citealt{Eisenstein:2011}) and utilized a new infrared spectrograph
to obtain moderate-resolution, H-band spectra for over 160,000 stars throughout
the Galaxy.
From these spectra, high-precision radial velocities, chemical abundances, and
stellar parameters have been derived and released as a part of DR13
(\citealt{Holtzman:2015,Nidever:2015}).

As a part of the observing strategy of \apogee, most stars are observed
multiple times and binned by day into ``visit'' spectra.
Though a typical star is only observed a few times, (1) at least one pair of
visits are separated by one month or longer, and (2) thousands of stars have
been observed more than 10 times (for a more detailed look at the cadence and
number of visits for \apogee\ stars, see \figname~1 in \citealt{Troup:2016}).
Radial velocities (and stellar parameters) are derived for each of the visit
spectra, affording a sparse and sporadic time-domain sampling of the radial
velocity variations of most stars in the survey.
This time-domain information was recently used to identify a sample of
candidate stellar and substellar companions to \apogee\ stars
(\citealt{Troup:2016}).

This search was conducted after data quality and data \emph{quantity} cuts that
were designed to keep the number of data points larger than the number of
parameters in the model.
In their case, the model parameters are six Keplerian orbital parameters plus a
long-term (linear) velocity trend (seven in total).
Under this criterion, stars with fewer than eight visits were eliminated from further
consideration, leaving $\approx$15,000 stars.
For each of the remaining stars, the radial velocity curves were searched for
significant periods, which were then used to initialize $\chi^2$ fits of a
Keplerian orbit using a custom least-squares fitter (\acronym{MPRVFIT};
\citealt{De-Lee:2013}).
In cases where multiple significant periods were found, the parameters obtained
from the fit with the best modified $\chi^2$ value were retained (modulo a
number of other considerations described in \sectionname~3.3.4 of
\citealt{Troup:2016}).

The complexity of this pipeline and logic needed to identify a single optimal
set of orbital parameters from a set of solutions highlights the fact that the
likelihood function for a Keplerian orbit model is generically multi-modal.
When there are few data points or poor phase coverage this is especially true.
While useful for searching for new candidate binaries, such a pipeline does not
easily fit within hierarchical probabilistic modeling of the population of
companions.

\subsection{Experiment~\arabic{expcounter}: Validation with simulated data}
\label{sec:validation}
\stepcounter{expcounter}

As a first demonstration, we generate fake radial velocity observations (with
uncertainties) that are consistent with our assumptions
(\sectionname~\ref{sec:method}), then sample from the posterior pdf over orbital
parameters with \samplername.
The eccentricity, period, and velocity semi-amplitude are chosen to be broadly
consistent with the typical sub-stellar companion found in recent analysis of
the \apogee\ data (\citealt{Troup:2016}), the angle parameters ($\omega$,
$\phi_0$) are sampled from a uniform distribution, and the barycenter velocity
is drawn from a zero-mean Gaussian with variance $\sigma^2 = (30~\kms)^2$.
The values of the parameters used for the simulated data (i.e. the truth) are:
$P = 103.71~{\rm day}$, $e = 0.313$, $\omega = 68.95^\circ$,
$\phi_0 = 223.96^\circ$, $K = 8.134~\kms$, $v_0 = 42.98~\kms$.
We randomly sample five observation epochs uniformly over the interval $(0,1095)~{\rm day}$
imagining a 3-year survey with no observing strategy and arbitrarily set the
survey start date (in barycentric MJD) to be 55555.
The radial velocity measurement uncertainties are drawn from a uniform
distribution over the interval $(0.1, 0.2)~\kms$, motivated by current \apogee\
radial velocity uncertainties.

We perform two samplings as a validation of \samplername:
(a) we fix the jitter parameter $s^2 = 0$, i.e. we assume the uncertainties are
known perfectly and there is no intrinsic scatter, and (b) we sample over the
jitter parameter as well, setting $(\mu_s,\sigma^2_s) = (0,8)$ (note that these
are dimensionless because they set the scale of a Gaussian in $\ln s^2$).
We start by generating $J=2^{28}$ prior samples over the nonlinear parameters
with a period domain of $(P_{\rm min}, P_{\rm max}) = (16, 8192)~{\rm day}$ (see
\sectionname~\ref{sec:method}) and use these same prior samples for both runs.
For (a), 678 samples pass our rejection sampling step, and for (b), 580 samples
survive.

\begin{figure*}[p]
\begin{center}
\includegraphics[width=\textwidth]{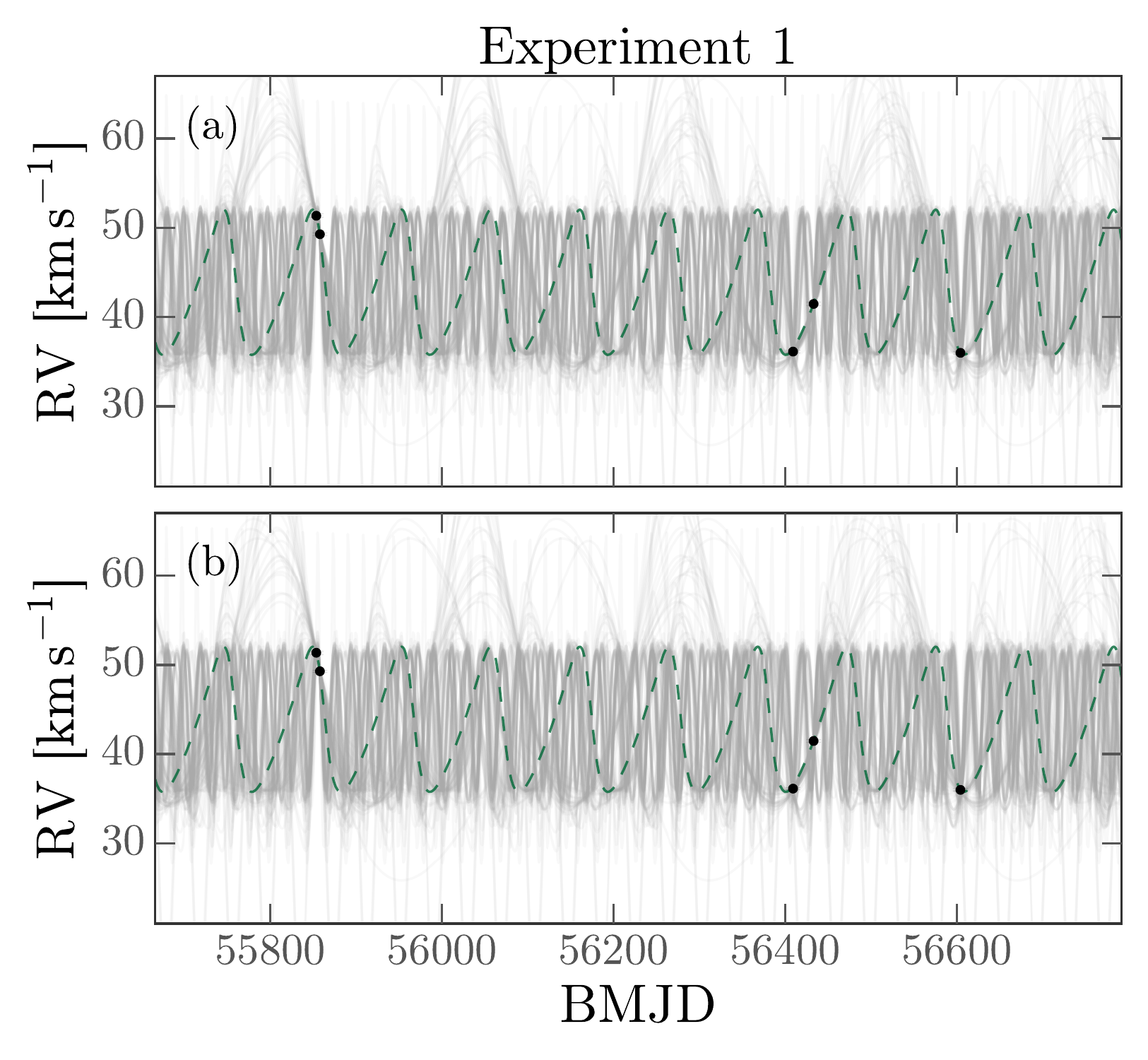}
\end{center}
\caption{%
In both panels, black points show five simulated radial velocity measurements, plotted
against Barycentric Modified Julian Date (BMJD) of the observations. The dashed
green line shows the true, input orbit from which the radial velocity measurements
were drawn.
Uncertainties on the data points are much smaller than the marker size.
Gray curves show orbits computed from 128 samples from, in the top panel, the
posterior pdf with the jitter held fixed at $s^2 = 0$ and, in the bottom panel,
the posterior pdf including jitter as a free parameter.
Using \samplername\ several qualitatively different orbital solutions are found over a range of
eccentricities and periods for each case, including the correct orbit.
This is simply a validation of the method.
\label{fig:validation-rv}}
\end{figure*}

\begin{figure*}[p]
\begin{center}
\includegraphics[width=\textwidth]{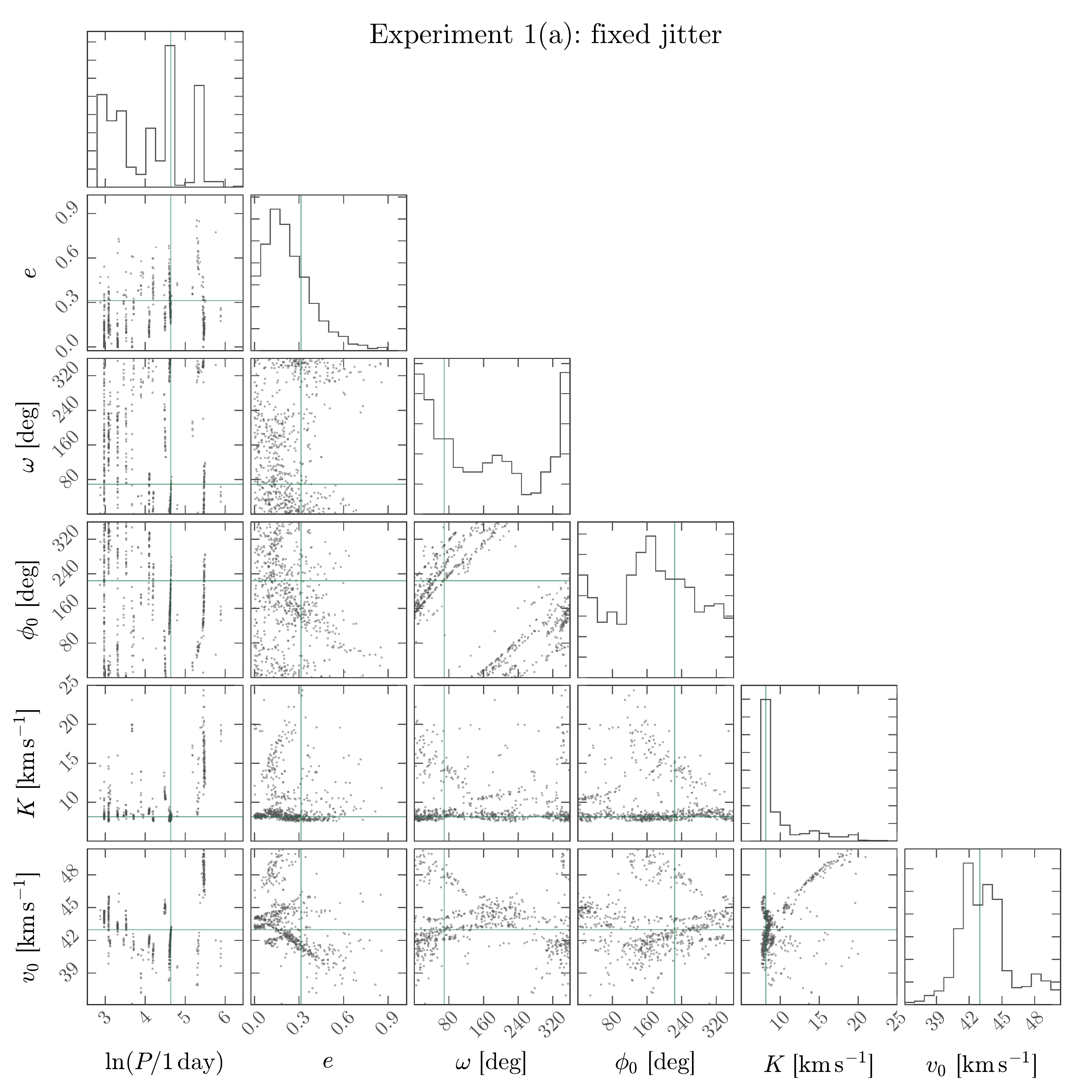}
\end{center}
\caption{%
Projections of the 678 surviving posterior samples when considering the five velocity measurements from
Fig.\ref{fig:validation-rv} with fixed jitter (gray
points).
The values used to generate the input orbit are shown as green cross-hairs.
\label{fig:validation-corner-a}}
\end{figure*}

\begin{figure*}[p]
\begin{center}
\includegraphics[width=\textwidth]{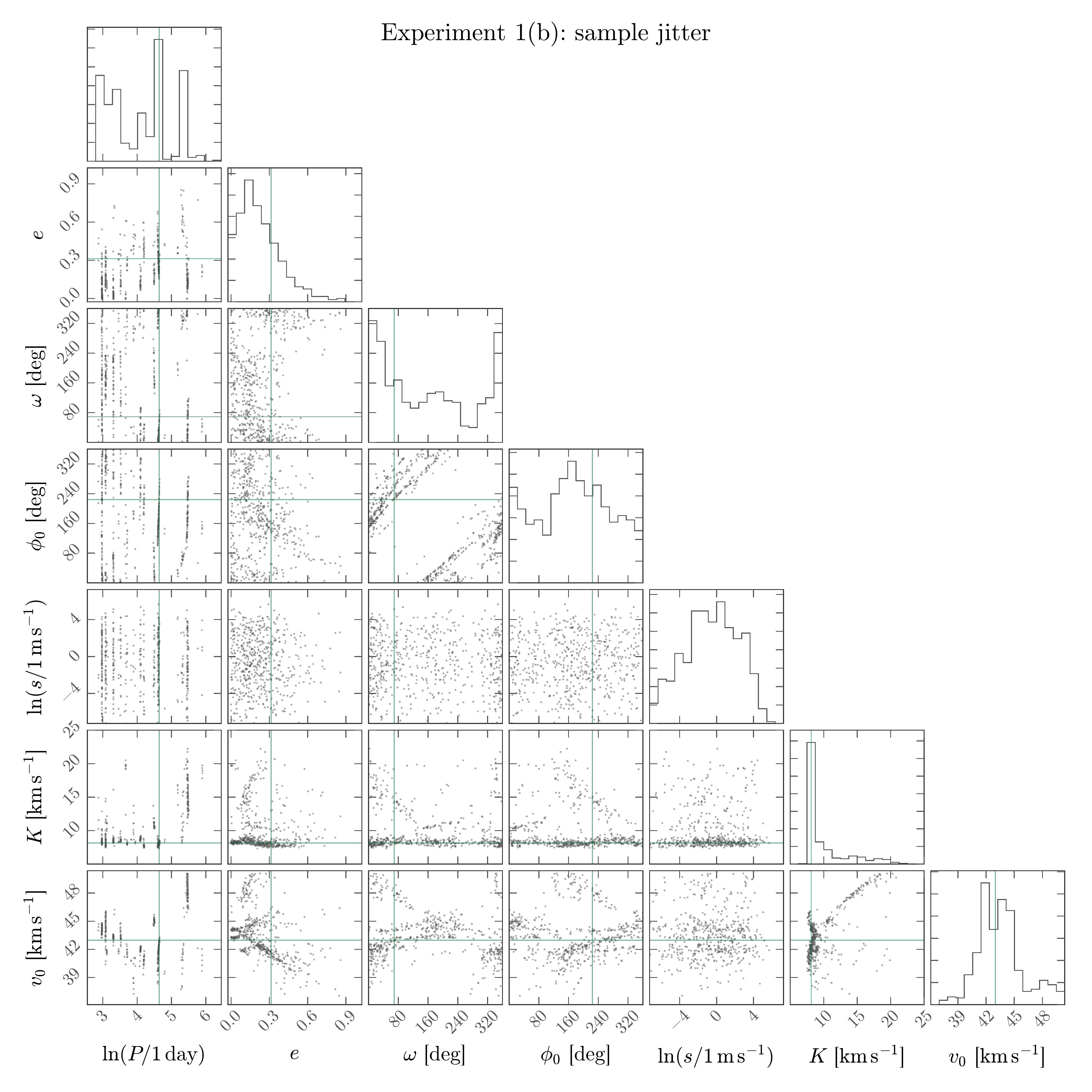}
\end{center}
\caption{%
Same as \figname~\ref{fig:validation-corner-a} but for the 580 surviving posterior
samples when also sampling over an unknown jitter parameter, $s^2$.
\label{fig:validation-corner-b}}
\end{figure*}

\figname~\ref{fig:validation-rv} shows the simulated data (black points) along with
the true orbit (dashed, green line) and orbits computed from samples from the
posterior pdf (gray lines) for the sampling with fixed jitter (top panel) and
the sampling including jitter as a non-linear parameter (bottom panel).
Because we use the same prior samples in each, these look almost identical.
\figname s~\ref{fig:validation-corner-a}--\ref{fig:validation-corner-b} show all
projections of the posterior samples (gray points) and the true, input
parameters (green lines and markers) for the case with fixed jitter and the case
where jitter is treated as a free parameter.
The surviving samples in each case look very similar, as expected.
The typical uncertainties for these simulated data are $\sigma_v \approx
0.15~\kms$; for values of $\ln s \lesssim 3$ (where most are concentrated), this
extra jitter is negligible compared to the uncertainties.
Note that samples above $\ln s \approx 5$ are mostly rejected, indicating that,
as constructed, the uncertainties are purely Gaussian and known.

In both cases, the modes in period-space are narrow with a variety of
separations, as can be seen in the radial-velocity curves plotted in the top
panel.
For small numbers of precise observations with poor phase-coverage, the
posterior pdf over orbital parameters is extremely complex and structured, but
we are still able to generate samples using \samplername\ that capture the
complexity.
It is obvious that these multimodal samples are essentially useless when
trying to understand one particular system at hand. Yet, they rule out almost
all of the parameter space encompassed by the prior distribution:
even a few radial velocity measurements are manifestly very informative.

\subsection{Experiment~\arabic{expcounter}: Underestimated uncertainties}
\label{sec:undunc}
\stepcounter{expcounter}
Besides the number and spacing of the observation epochs, the
precision of the individual radial velocity measurements matters.
Less precise data, compared to the radial velocity amplitude,
will admit more qualitatively different orbital solutions.
The structure and complexity of the posterior pdf (at fixed jitter) will
therefore depend strongly on knowing the measurement uncertainties and
the intrinsic velocity variability of the system.
We illustrate this using another simulated radial velocity curve with
lower signal-to-noise. We show that for underestimated uncertainties,
the posterior pdf over orbital parameters can look well-constrained, but may be
discrepant with the true orbital parameters.
Finally, we show that by simultaneously sampling over an unknown extra variance in
the data (the jitter), we can account for additional, unaccounted sources of noise.

For this experiment, we use the following parameter values to generate the
simulated data: $P = 103.71~{\rm day}$, $e = 0.313$, $\omega = 250.73^\circ$,
$\phi_0 = 103.01^\circ$, $K = 1.134~\kms$, $v_0 = 8.489~\kms$.
We again uniformly sample five observation epochs over the interval $(0,1095)~{\rm
day}$ and use radial velocity measurement uncertainties drawn from a uniform
distribution over the interval $(0.1, 0.2)~\kms$ (the median true uncertainty is
$\ln \left(\frac{\sigma_v}{\mps}\right) \approx 5$).

When running \samplername\ for this experiment, we consider three cases:
(a) the uncertainties are (correctly) known and jitter is fixed and ignored ($s^2=0$), (b)
the uncertainties are underestimated by a factor of 8 and jitter is fixed and
ignored ($s^2=0$), and (c) the uncertainties are underestimated by a factor of 8
and we treat the jitter as a free parameter.
For each case, we again generate $2^{28}$ prior samples over the nonlinear
parameters with a period domain of $(P_{\rm min}, P_{\rm max}) = (16, 8192)~{\rm
day}$ and re-use these prior samples for each case.
In case (c), we generate samples in the jitter by setting
$(\mu_s,\sigma^2_s) = (10,1)$---here we are assuming that we have some suspicion
about the true magnitude of the uncertainties.

\begin{figure*}[p]
\begin{center}
\includegraphics[width=\textwidth]{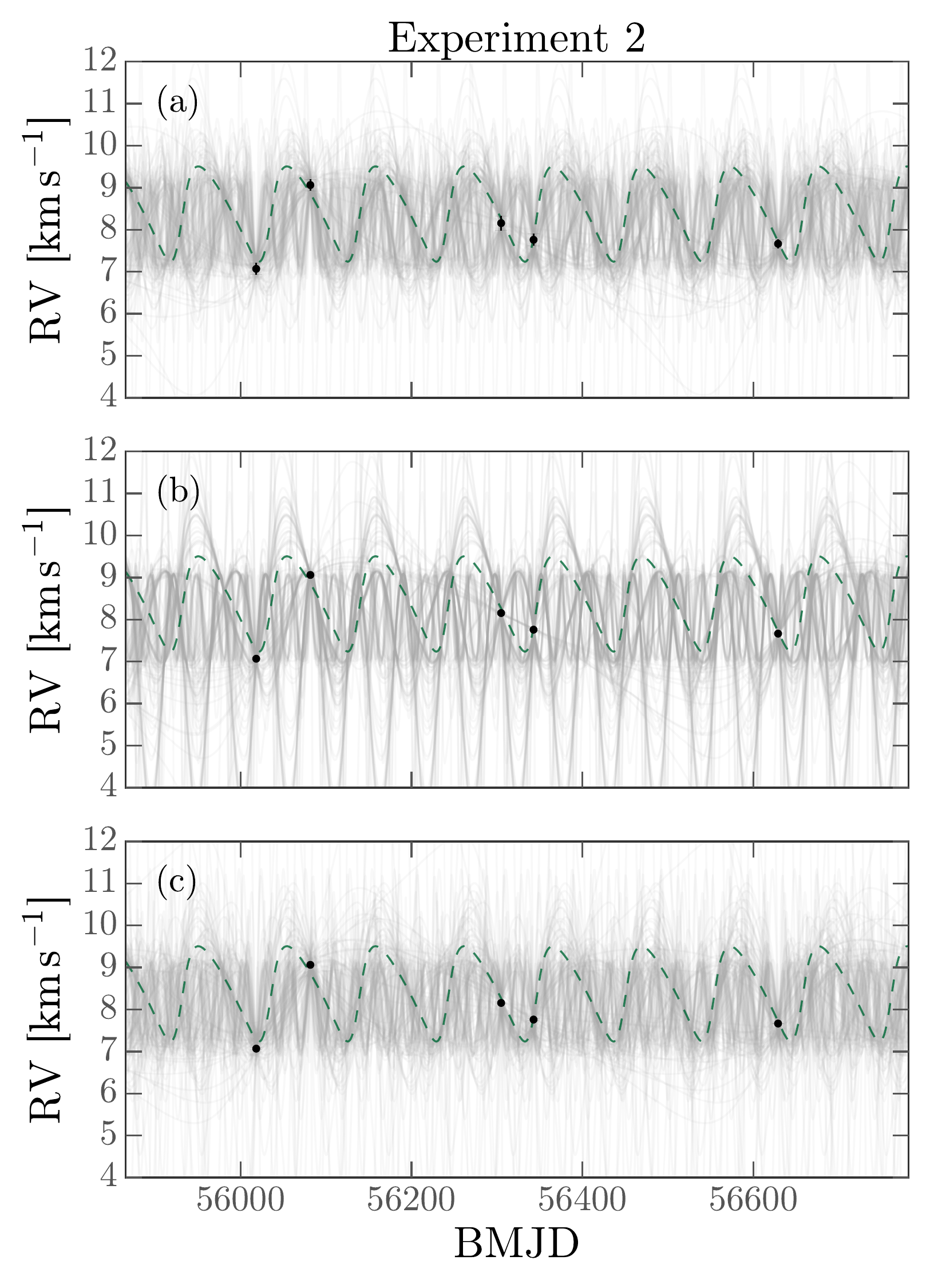}
\end{center}
\caption{%
Same as \figname~\ref{fig:validation-rv} but for the three cases in this experiment:
(a) known uncertainties and jitter fixed to $s^2 = 0$, (b) underestimated
uncertainties and jitter fixed to $s^2 = 0$, (c) underestimated uncertainties
and jitter allowed to vary.
\label{fig:undunc-rv}}
\end{figure*}

\begin{figure*}[p]
\begin{center}
\includegraphics[width=\textwidth]{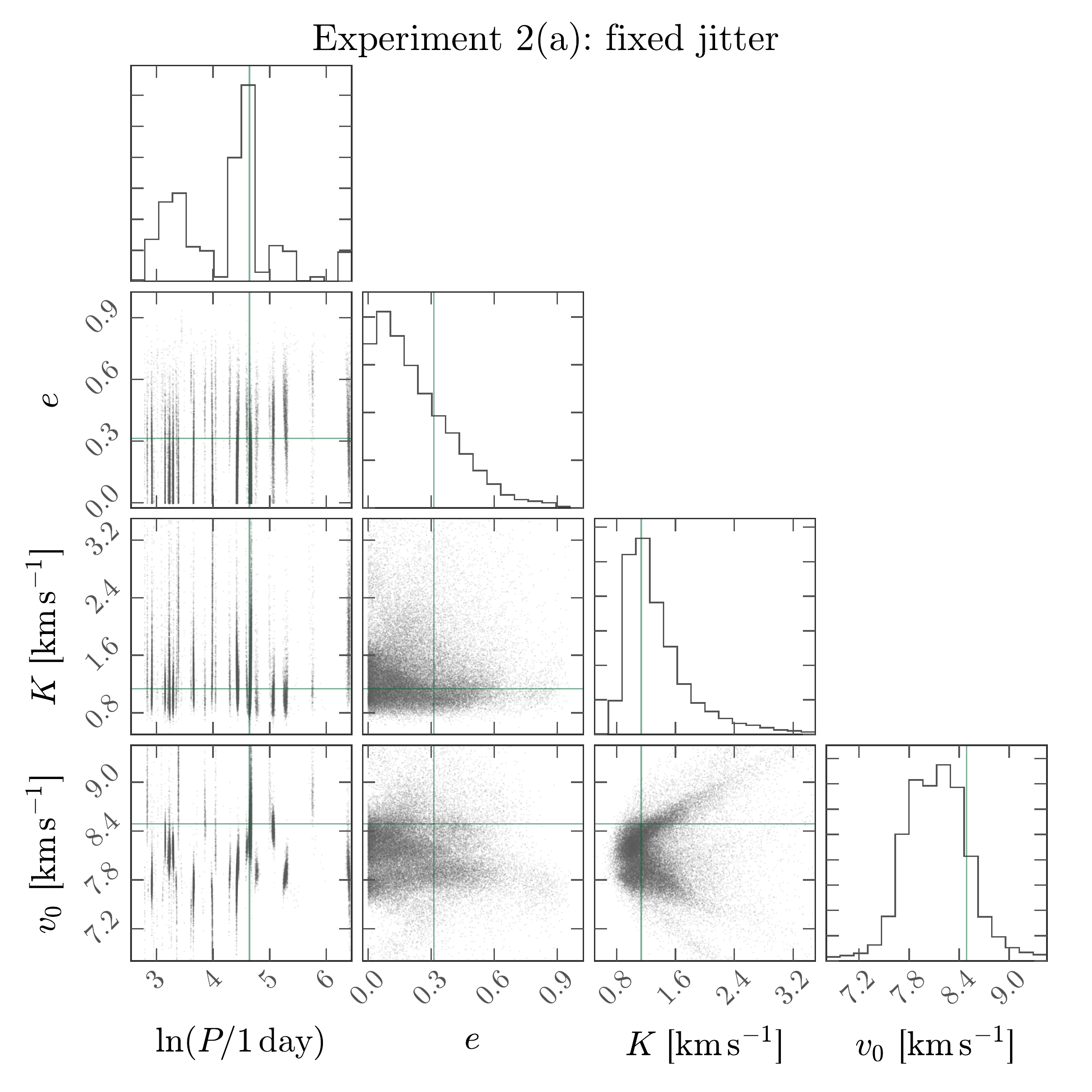}
\end{center}
\caption{%
Same as \figname~\ref{fig:validation-corner-a} but for case (a) in Experiment 2, where the
uncertainties are known and jitter is fixed to $s^2 = 0$.
46330 samples survive the rejection step.
Henceforth we will drop the angular parameters $\omega$ and $\phi_0$ from the
corner plots to conserve space.
\label{fig:undunc-corner-a}}
\end{figure*}

\begin{figure*}[p]
\begin{center}
\includegraphics[width=\textwidth]{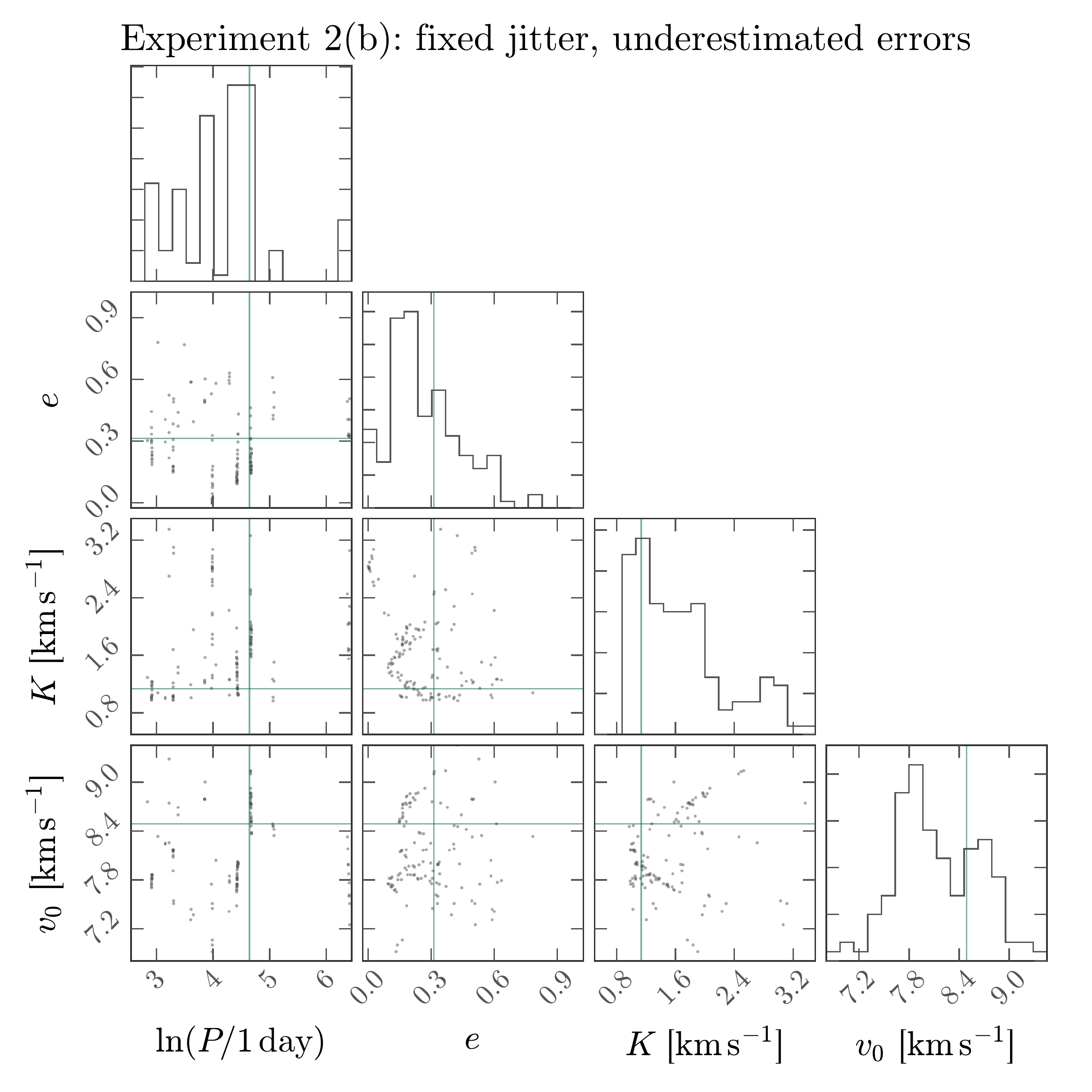}
\end{center}
\caption{%
Same as \figname~\ref{fig:undunc-corner-a} but for case (b) in Experiment 2, where the
uncertainties are underestimated by a factor of 8 and jitter is fixed to
$s^2 = 0$.
146 samples survive the rejection step.
\label{fig:undunc-corner-b}}
\end{figure*}

\begin{figure*}[p]
\begin{center}
\includegraphics[width=\textwidth]{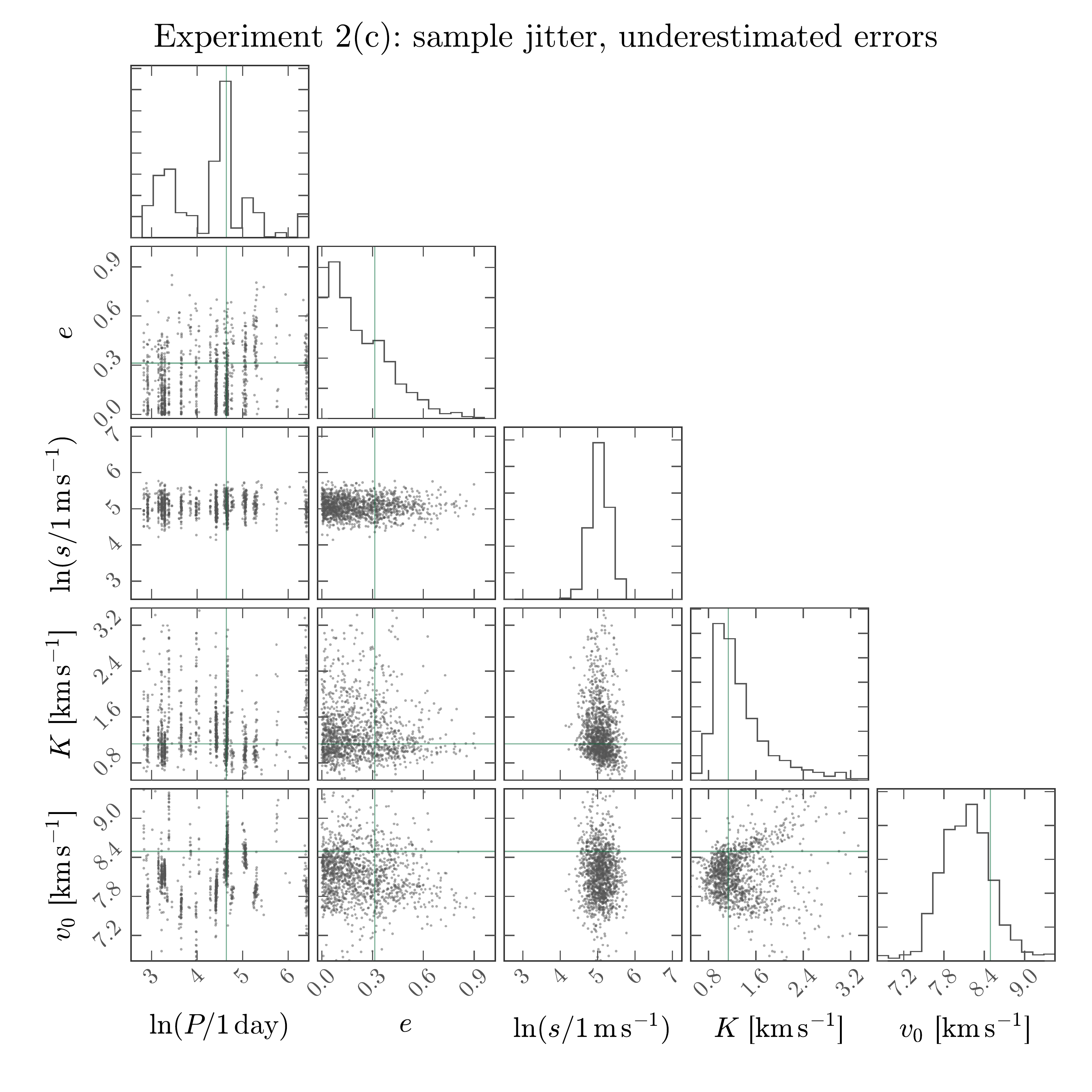}
\end{center}
\caption{%
Same as \figname~\ref{fig:undunc-corner-a} but for case (c) in Experiment 2, where the
uncertainties are underestimated by a factor of 8 and the jitter is sampled as
a non-linear parameter.
1323 samples survive the rejection step.
\label{fig:undunc-corner-c}}
\end{figure*}

\figname~\ref{fig:undunc-rv} shows the simulated data and presumed uncertainties (black
points), the true orbit (dashed, green line), and orbits computed from samples
from the posterior pdf (gray lines) for the three different samplings.
In all cases, a wide variety of orbit solutions is permitted by the data.
Note that far fewer modes are present in the middle panel (case (b)), when the
uncertainties are underestimated.
\figname s~\ref{fig:undunc-corner-a}--\ref{fig:undunc-corner-c} show the
corresponding corner plots of the parameter pdf's for these three cases.
For this data set, when the uncertainties are correctly known, the posterior pdf
is highly multi-modal (case (a)).
When the uncertainties are underestimated, the posterior pdf has fewer modes,
but the true orbital parameters do not appear consistent with any of the
strongest modes (case (b)).
When the uncertainties are (severely) underestimated but the jitter is allowed
to vary (case (c)), the model prefers solutions with a finite jitter comparable
to the input true uncertainties ($\ln \left(\frac{\sigma_v}{\mps}\right) \approx
5$).
We have therefore shown that \samplername\ is useful even when uncertainties are
underestimated or the intrinsic velocity variability of a system is unknown.

\subsection{Experiment~\arabic{expcounter}: Varying the number of data points}
\label{sec:numpts}
\stepcounter{expcounter}

When the phase coverage of the radial velocity observations is good and the
number of observation epochs is large, the posterior pdf over orbital parameters
effectively becomes unimodal.
Under these conditions, \samplername\ is of course a very inefficient sampler for this
problem and will return very few samples (as few as one).
As we have seen in the previous experiments, when the number of data points is
small or the uncertainties are large, the posterior pdf is generally
multi-modal.
In this experiment we explore the dependence of the posterior pdf's complexity
on the number of observation epochs by generating radial velocity curves with
initially 11 epochs.
We use the following parameter values to generate the simulated data: $P =
103.71~{\rm day}$, $e = 0.313$, $\omega = 134.83^\circ$, $\phi_0 =
342.26^\circ$, $K = 4.227~\kms$, $v_0 = 19.431~\kms$.
After running \samplername\ with the full 11 observations, we successively
remove 2 data points and re-run the sampling until we are left with three
observations epochs as input data (a total of five consecutive runs).

\begin{figure*}[p]
\begin{center}
\includegraphics[width=\textwidth]{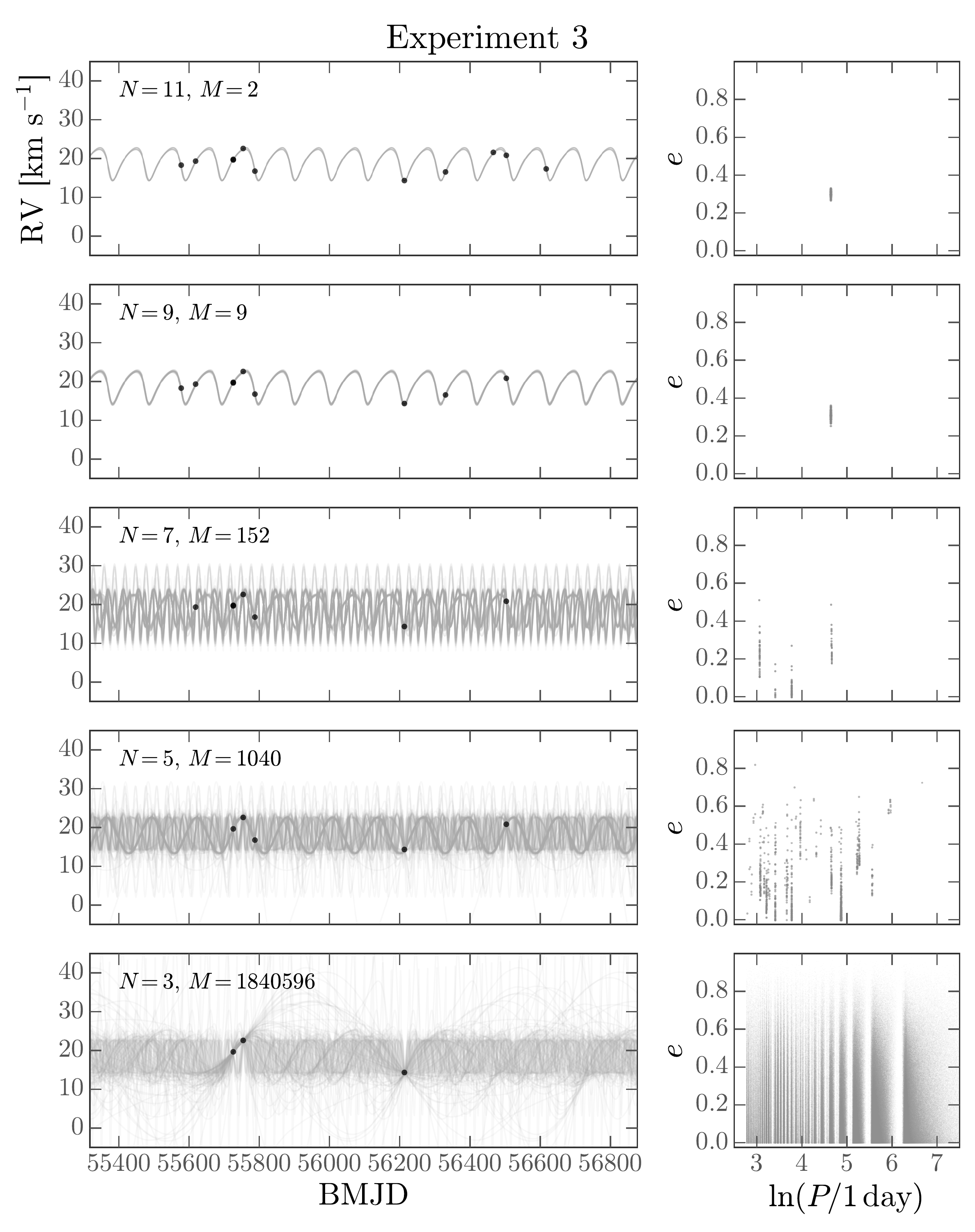}
\end{center}
\caption{%
{\sl Left panels}: Analogous to the top panel of
\figname~\ref{fig:validation-rv}, black points show simulated radial velocity
measurements used to generate samples from the posterior pdf over orbital
parameters sub-sampled from the full set of simulated data.
Gray curves show 128 orbits computed from posterior samples, either from
\samplername or from switching to \emcee\ to continue sampling until 128 samples
are returned (as occurs for the top two panels).
The number of data points, $N$, and the number of samples that pass the
rejection sampling step of \samplername, $M$, are shown on each panel.
Note that two of the (randomly chosen) observation times are so close that the
markers overlap.
{\sl Right panels}: A single projection of the posterior samples in each case
showing the log-period, $\ln P$, and eccentricity, $e$.
The structure of the posterior pdf when the number of data points is small is
highly complex but still much more compact than our prior beliefs.
\label{fig:numpts-rv}}
\end{figure*}

Specifically, we again generate $2^{28}$ prior samples over the nonlinear
parameters with a period domain of $(P_{\rm min}, P_{\rm max}) = (16, 8192)~{\rm
days}$ and re-use these prior samples for each sub-sampling of the data.
We fix the jitter to $s^2 = 0$ and assume that the uncertainties are known.
\figname~\ref{fig:numpts-rv} shows the simulated data and orbits computed from
posterior samplings.
Starting from the top of \figname~\ref{fig:numpts-rv} with the full set of 11
data points, each panel below has two epochs fewer than the previous.
The data used for the pdf sampling shown in a given panel are plotted as black
circles and the number of data points used in each panel, $N$, are indicated.
As described in \sectionname~\ref{sec:method}, when the number of surviving
samples $M < M_{\rm min}=128$ after rejection sampling, we either (1) initialize
\emcee\ using the remaining sample(s) if the periods of the surviving sample(s)
are sufficiently close, or (2) re-run \samplername\ with a new set of prior
samples until we have at least $M_{\rm min}$ samples from the posterior pdf.
In all panels, 128 orbits computed from the posterior samples are shown.

The structure in the posterior samples in one projection of the posterior pdf
(period and eccentricity) is
shown in the right panels of \figname~\ref{fig:numpts-rv}.
For the cases with nine and 11 data points, the posterior pdf appears to be
unimodal.
Multiple modes first appear when $N=7$ and the posterior pdf become more
structured in further sub-samplings of the data until ultimately forming a
harmonic series of modes in the final case of $N=3$.
It is worth emphasizing that even for the case with 3 data points, $<1\%$ of the
prior samples pass the rejection step:
even 3 radial velocity observations are informative!

\subsection{Experiment~\arabic{expcounter}: Real data for a known binary}
\label{sec:apogee}
\stepcounter{expcounter}

\begin{figure*}[p]
\begin{center}
\includegraphics[width=\textwidth]{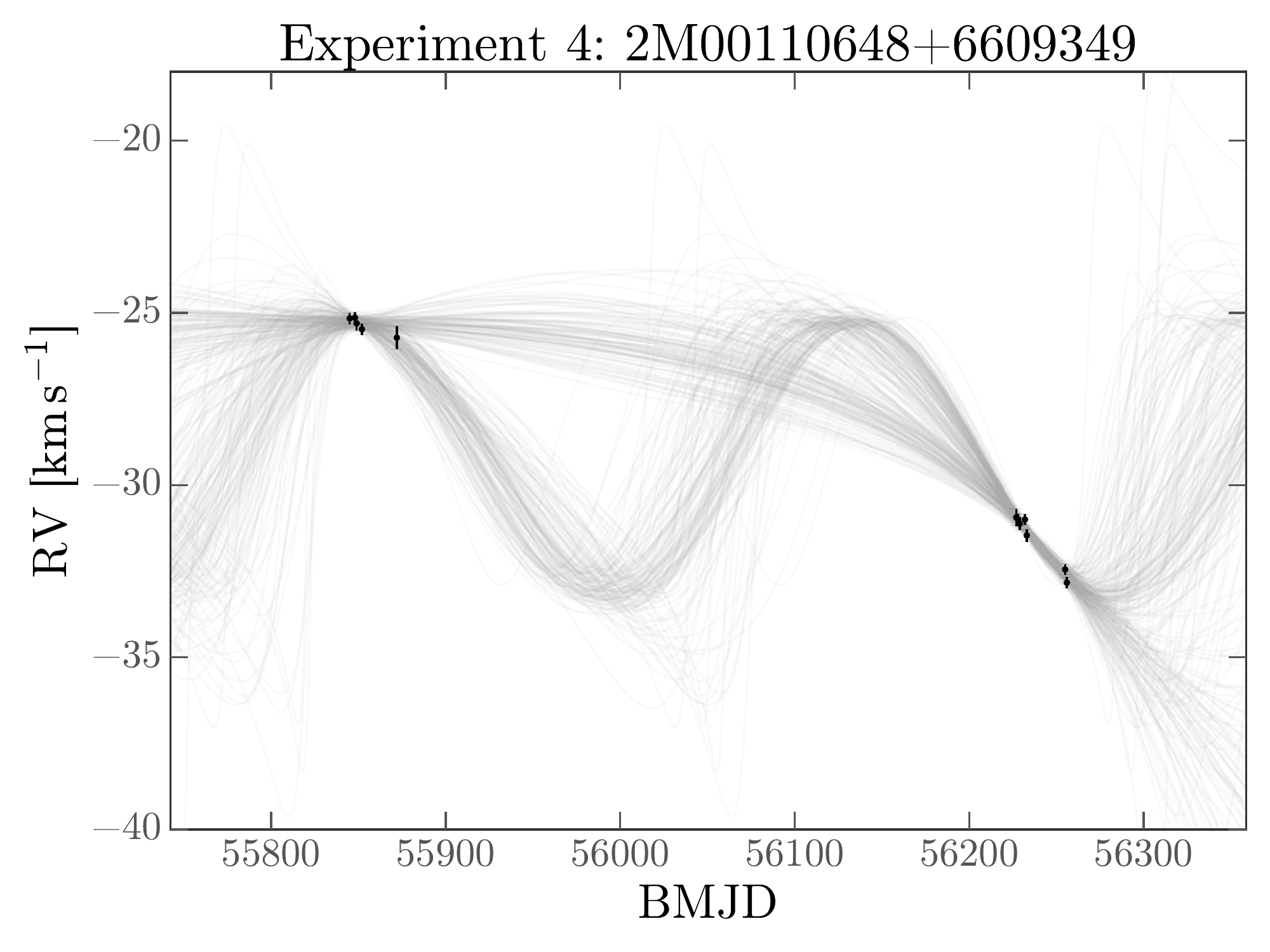}
\end{center}
\caption{%
Black points and uncertainties show \apogee\ radial velocity measurements for
the target 2M00110648+6609349.
Gray curves show orbits computed from 256 samples from the posterior pdf
generated with \samplername\ with jitter as a free parameter.
Several qualitatively different orbital solutions are found for this source.
\label{fig:apogee-rv}}
\end{figure*}

For a more realistic application of \samplername, we choose an \apogee\ target
with a previously identified companion (2M00110648+6609349) but with few radial
velocity measurements (\citealt{Troup:2016}).
\figname~\ref{fig:apogee-rv} shows radial velocity data for the star (black
points). 
Similar to Experiment 1, these data are sparse and have poor phase coverage.
However, this epoch sampling is quite different and is representative of realistic
survey design choices.

We again generate $2^{28}$ prior samples over the nonlinear parameters with a
period domain of $(P_{\rm min}, P_{\rm max}) = (16, 8192)~{\rm day}$ and with
$(\mu_s,\sigma^2_s) = (10.5,1)$, of which 22,313 samples pass our rejection
sampling step.
Over-plotted as gray lines on \figname~\ref{fig:apogee-rv} are 256 orbits
computed from these samples.
Already from visualizing these orbits, it becomes clear that there are at least a few
distinct period modes, and a wide variety of eccentricities, represented in the
posterior sampling.

\begin{figure*}[p]
\begin{center}
\includegraphics[width=\textwidth]{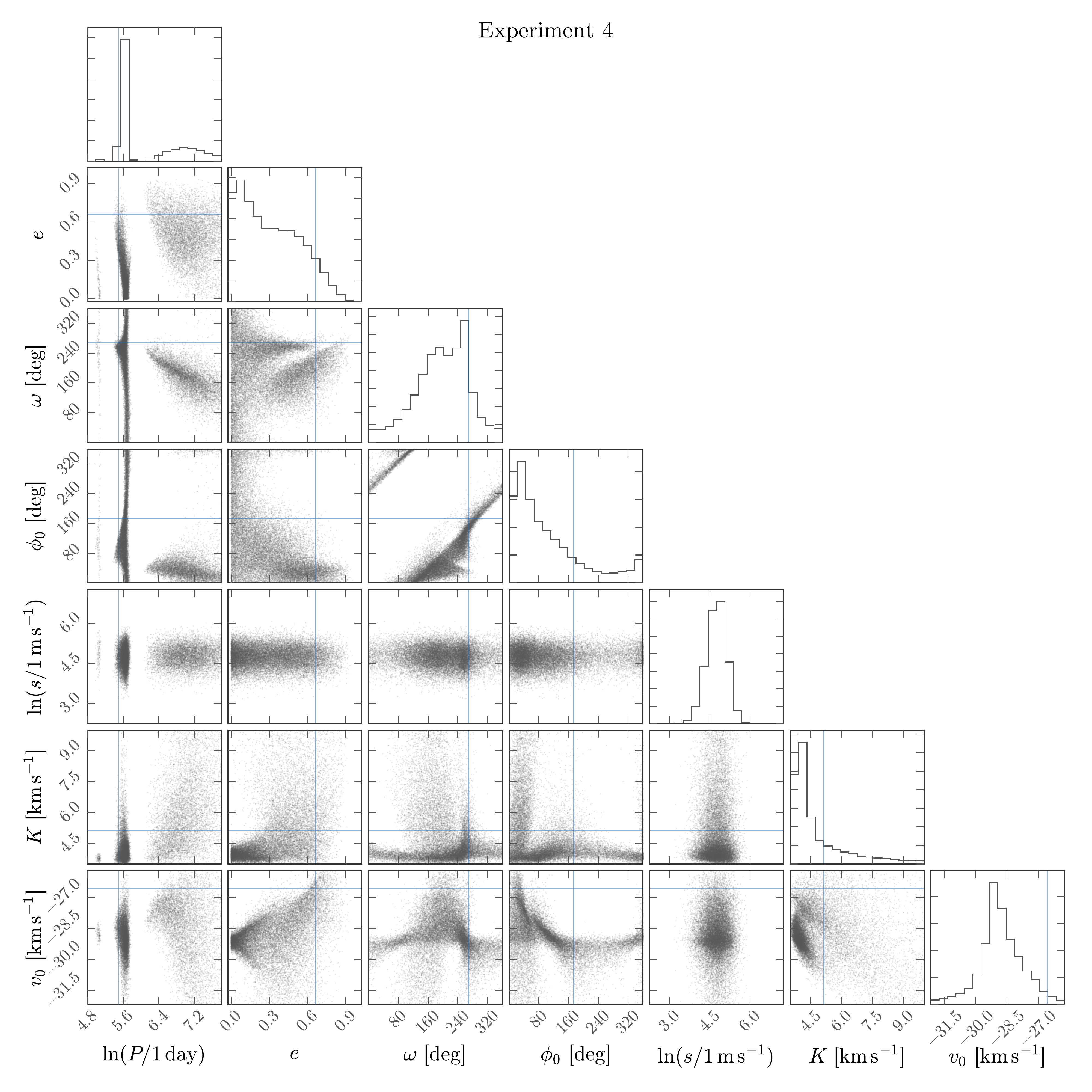}
\end{center}
\caption{%
All projections of the 22,313 surviving posterior samples (gray points)
for 2M00110648+6609349 with
previously found orbital parameter values shown as the blue cross-hairs
(\citealt{Troup:2016}).
\label{fig:apogee-corner}}
\end{figure*}

\figname~\ref{fig:apogee-corner} shows projections of all posterior samples in
different parameter combinations.
Here it is clear there are at least three period modes: the dominant mode at $P
\approx 300~{\rm day}$ is broadly consistent (but not coincident) with the previously measured period
(\citealt{Troup:2016}); but other modes are clearly present at shorter periods
with low eccentricity and longer periods with higher eccentricity.
Interestingly, we also find that the model prefers having a finite jitter
around $s \approx 116~\mps$, which would imply that the effective radial velocity
uncertainties might be underestimated by a factor of $\approx 2$.

\subsection{Experiment \arabic{expcounter}: Prospects for observation planning}
\label{sec:obsplan}
\stepcounter{expcounter}

A noticeable difference between the $N=5$ and $N=7$ panels in
\figname~\ref{fig:numpts-rv} is that the posterior pdf collapses significantly
between these cases (from $\gtrsim$20 period modes to 3):
This implies that the two added observations are extremely informative.
Inverting this idea, it also suggests that we can use \samplername\ to (1)
predict the observation time that maximally collapses the posterior pdf for a
previously measured source, and (2) for an expected population of sources, we
can identify the optimal sampling pattern to maximize discovery or
characterization of the sources.
We will explore these ideas in detail in future work, but here we simply show
that the timing of subsequent observations can lead to very different
structure in the posterior samples.

\begin{figure*}[p]
\begin{center}
\includegraphics[width=\textwidth]{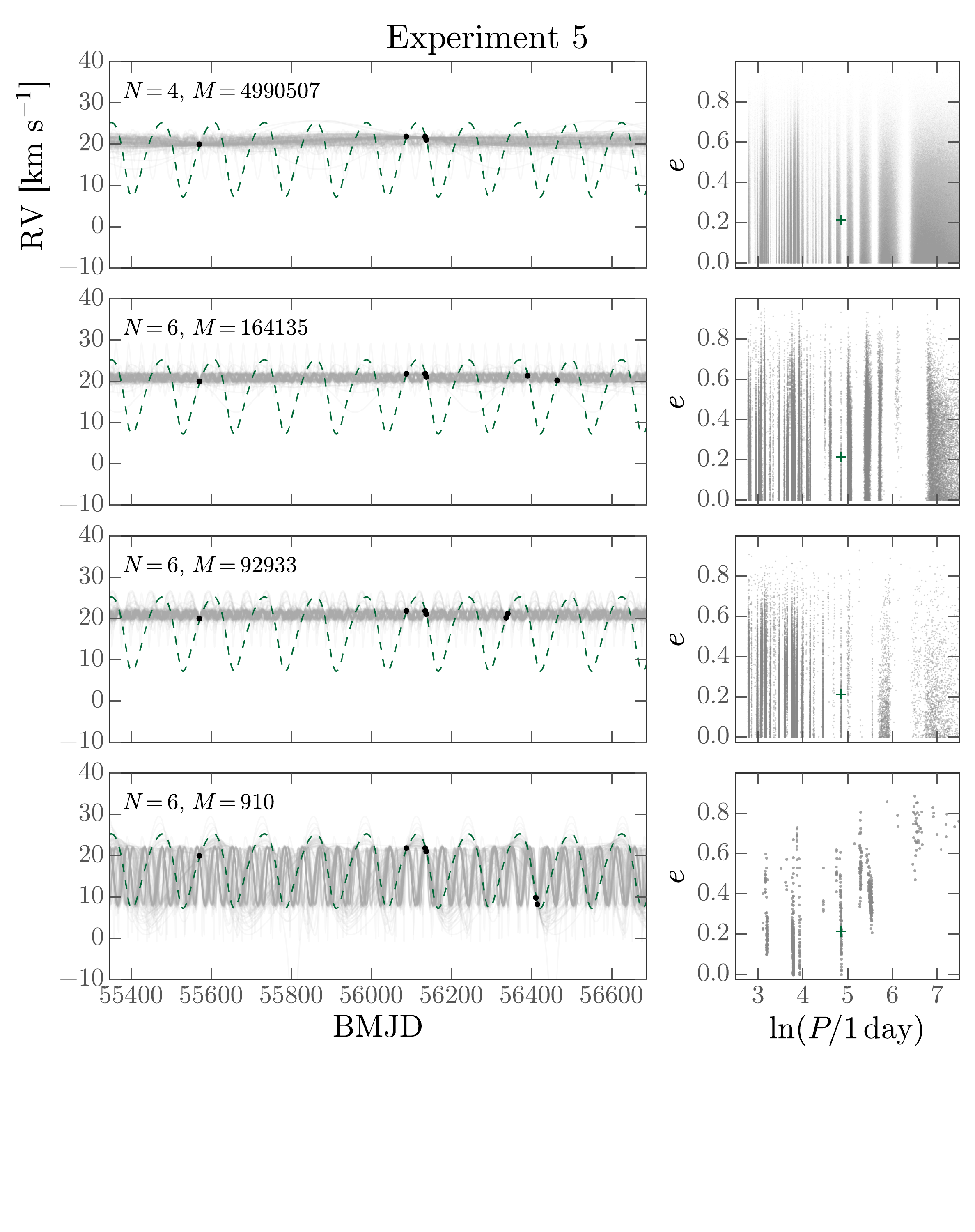}
\end{center}
\caption{%
Similar to \figname~\ref{fig:numpts-rv} but now varying the timing of new
simulated observations relative to the data in the top-left panel.
The top row has four simulated observations and all of the rows below the top
row have two additional observations (six) placed at different times but with
fixed spacing in phase.
The number of data points $N$ and number of samples that survive the rejection
sampling $M$ are shown on each panel.
\label{fig:obsplan-rv}}
\end{figure*}

We again simulate a data set of four noisy radial velocity measurements, shown
as black points in the top-left panel of \figname~\ref{fig:obsplan-rv}.
We use the following parameter values to generate the simulated data:
$P = 127.31~{\rm day}$, $e = 0.213$, $\omega = 137.23^\circ$, $\phi_0 =
36.23^\circ$, $K = 8.996~\kms$, $v_0 = 17.643~\kms$.
Uncertainties were chosen to match the \apogee\ data ($\sigma_v \approx
0.2~\kms$) and are shown as error bars, but they are often comparable to or
smaller than the marker size.
The top-right panel shows posterior samples produced with \samplername\ again in the
space of log-period, $\ln P$, and eccentricity, $e$.
The three lower rows all have six observation epochs: the same four from the top row,
but now with two additional observations spaced, in phase, by $\frac{\Delta
\phi}{2 \pi} = 0.04$ but with a different starting phase for the new
observations.
As is shown by the right panels, the observation times of the new observations
can greatly effect the compactness of the posterior pdf.
In particular, the placement of the observations in the bottom row of panels
rules out most of the long-period modes from the top panels, \emph{and} many of
the short-period modes, whereas for the middle two cases the new data are not as
informative.

\section{Discussion} \label{sec:discussion}

We have built a Monte Carlo sampler---\samplername---to draw samples from the
full posterior pdf over orbital parameters for single-companion systems,
given a set of multi-epoch, single-line radial velocity measurements.
\samplername\ has important properties that differ from other sampling methods:
(1) It produces \resp{independent} samplings even when the likelihood (and hence
posterior pdf) is highly multi-modal; (2) the method is based on Simple Monte
Carlo,  in some sense a pure brute-force method, which parallelizes trivially;
and (3) the samplings are guaranteed to be correct under the sensible
assumptions presented here, without the need for convergence or other diagnostic
checks.
If the pdf is effectively unimodal, \samplername\ tells us that the solution is
unique.
If the pdf is multi-modal, \samplername\ captures all relevant different
solutions.

Our experiments show that \samplername\ can be used for discovery and
characterization of stellar binaries or exoplanets, even with the presence of
unrecognized or unaccounted noise contributions.
However, for exoplanets, we emphasize that while \samplername\ could in
principle be used for any exoplanet system, the simplicity of our noise model
and single-companion assumption strongly suggest that it will be most useful in
the study of massive exoplanets.

Perhaps the primary innovation \samplername\ brings is a separation of the
parameters into linear and nonlinear subsets.
The brute force sampling is only required in the nonlinear subspace, aiding
computational feasibility.
We further capitalize on the problem structure by identifying effectively
unimodal and effectively multi-modal posterior pdfs using the minimum possible
width of a likelihood peak in the period direction, given the time sampling.

Interestingly, as we show in \sectionname~\ref{sec:numpts}, even very sparse
samplings of the radial-velocity history of a star provide highly
informative posterior pdfs.  The bottom-right panel in
\figname~\ref{fig:numpts-rv} shows a highly multi-modal posterior pdf.
Nonetheless, the vast majority of prior pdf samples have been eliminated, and
only a tiny subset of periods, eccentricities, and amplitudes are consistent
with the data.
These posterior pdfs may look bewilderingly complex, but they can contribute
extremely valuable information to any hierarchical inference, or
provide a very informative prior pdf for further observing campaigns.

Indeed, \samplername\  can be used to generate inputs for a hierarchical inference.
In previous work (\citealt{Hogg:2010, Foreman-Mackey:2014}) we have shown
that posterior samplings under an interim prior can be importance-sampled
with a hierarchical inference to generate posterior beliefs about the
full population.
These hierarchical inferences are the only population inferences
that properly propagate non-trivial uncertainties at the
individual-system level to the conclusions at the population level \resp{(see
\citealt{Mandel:2011, Strader:2004, Brewer:2013, Brewer:2014} for other examples
of hierarchical inference in astronomy)}.

In the experiments above we have used massive prior samplings, starting with
$2^{28}$ samples before rejection sampling.
When the data are sparse or have a low signal-to-noise (e.g., bottom panels in
\figname~\ref{fig:numpts-rv}), many prior samples pass the rejection-sampling
step:
If the goal is to learn about an individual system and the data are poor, many
fewer prior samples can be used to initialize \samplername.
\resp{In this limit, generating a set number of posterior samples is very fast
because of the easy parallelization of the likelihood calls.}
The same is true when the data are high-quality and the samples that pass the
rejection step will be used to initialize an MCMC sampling.
A large prior sampling is needed when (a) many posterior pdf samples are needed
for hierarchical inference, or (b) the data are of intermediate quality (the
exact definition of which is problem-specific).
\resp{\samplername\ is most valuable in the low-to-intermediate quality range,
especially when the samples will be used for hierarchical inference, when a
small but converged sampling is needed for many (thousands to millions) of
sources.}

\samplername\ should also be valuable in observation planning, or cadence
evaluation, or survey strategy:
As \sectionname~\ref{sec:obsplan} shows,
\samplername\ could be used to plan the times of next observations to maximize
their expected information content \resp{(see also
\citealt{Loredo:2004,Ford:2004})}.
That is beyond the scope of this \documentname\ and will be explored in future
work.

\samplername\ is based on a set of assumptions, itemized in
\sectionname~\ref{sec:method}.
The method delivers correct samplings when these specific assumptions hold.
Of course, these specific assumptions do \emph{not} hold sufficiently well!

Astrophysically, we know that a star's radial-velocity history
need not be set entirely by a single companion, with no other perturbers or
sources of radial-velocity signal.
Although the single-companion assumption is a severe assumption, it
is pretty-much required for the method to be tractable.
Of course, in reality, it is likely that many stars reside in higher-order
multiplets.
Sampling over orbital parameters even for two companions, however, is already
intractable as the non-linear parameter space jumps to eight-dimensional, and
(at least) ten-dimensional if there are companion--companion interactions.
This would be absolutely intractable to sample by brute-force Simple
Monte Carlo; our advice would be to switch to some kind of Markov
Chain method that deals as well as possible with multi-modal
posteriors, such as nested sampling (\citealt{Skilling:2004, Brewer:2009}).
This change would be associated with the loss of the simple
convergence criterion that the \resp{rejection sampling} provides:
If lots of samples survive the rejection step, the posterior has been sampled
\resp{independently}!
There is no comparably simple way of determining that any nested
sampling is converged.

That said, there is a simple $N$-body problem that \emph{can} be solved
tractably:
For systems with one short-period companion and $\geq1$ very long-period
companion(s), \samplername\ can be easily extended to include additional linear
parameters that allow long-period velocity trends that are, e.g., polynomial in
time; these additional parameters don't worsen the prior pdf sampling (which
happens over the nonlinear parameters only).
These extra linear parameters could alternatively include extra $v_0$ terms that
come in when, say, the data come from a set of different radial-velocity
programs with different calibrations (as is the case for the recent, impressive
Proxima b discovery; \citealt{Anglada-Escude:2016}).


At a crucial practical level, \resp{the assumption of Gaussian noise and
perfectly known noise variances is often violated.}
\resp{Here, introducing the jitter as an explicit fitting parameter should help
to mitigate \samplername's sensitivity to unknown noise sources.}
Nonetheless, presuming we understand the noise properties, may still be the most
problematic  assumption made by
\samplername :
Essentially all data sets show occasional outliers (catastrophic errors).
There is nothing we can do simply here, if we want to capitalize on treating the
linear and non-linear parameters separately.
To deal with very rare outliers (such that no star would be likely to suffer
from more than one), one possible modification would be to do all leave-one-out
samplings, take the union, and then importance sample the results using some
ratio of the mixture of leave-one-out Gaussian likelihoods to some more
realistic likelihood that involves an outlier model (as, for example, we suggest
in \citealt{Hogg:2010a}).
Such a modification to the method is beyond the scope of this
\documentname\, but not beyond the scope of our ambitions.

\acknowledgements
This project was started at AstroHackWeek 2016, organized by Kyle
Barbary (UCB) and Phil Marshall (SLAC) at the Berkeley Institute for
Data Science.
It is a pleasure to thank
  Megan Bedell (Chicago),
  Will Farr (Birmingham),
  Ben Weaver (NOAO),
  Josh Winn (Princeton),
  and the participants at AstroHackWeek 2016
for valuable discussions.
We thank the referee, Brendon Brewer, for extremely valuable feedback.

This research was partially supported by
  the \acronym{NSF} (grants \acronym{IIS-1124794}, \acronym{AST-1312863}, \acronym{AST-1517237}),
  \acronym{NASA} (grant \acronym{NNX12AI50G}),
  and the Moore-Sloan Data Science Environment at \acronym{NYU}.
The data analysis presented in this article was partially performed on
computational resources supported by the Princeton Institute for Computational
Science and Engineering (PICSciE) and the Office of Information Technology's
High Performance Computing Center and Visualization Laboratory at Princeton
University.
This work additionally used the Extreme Science and Engineering Discovery
Environment \citep[XSEDE;][]{Towns:2014}, which is supported by National
Science Foundation grant number ACI-1053575.

This project made use of \sdssiii\ data. Funding for \sdssiii\ has been
provided by the Alfred P. Sloan Foundation, the Participating Institutions, the
National Science Foundation, and the \acronym{U.S.} Department of Energy Office
of Science. The \sdssiii\ web site is \url{http://www.sdss3.org/}.

\sdssiii\ is managed by the Astrophysical Research Consortium for the
Participating Institutions of the \sdssiii\ Collaboration including the
University of Arizona, the Brazilian Participation Group, Brookhaven National
Laboratory, Carnegie Mellon University, University of Florida, the French
Participation Group, the German Participation Group, Harvard University, the
Instituto de Astrofisica de Canarias, the Michigan State/Notre
Dame/\acronym{JINA} Participation Group, Johns Hopkins University, Lawrence
Berkeley National Laboratory, Max Planck Institute for Astrophysics, Max Planck
Institute for Extraterrestrial Physics, New Mexico State University, New York
University, Ohio State University, Pennsylvania State University, University of
Portsmouth, Princeton University, the Spanish Participation Group, University
of Tokyo, University of Utah, Vanderbilt University, University of Virginia,
University of Washington, and Yale University.

\software{The code used in this project is available from
\url{https://github.com/adrn/thejoker} (\citealt{TheJoker:zenodo}).
The code used to run the experiments and generate the figures for this article
is available from \url{https://github.com/adrn/thejoker-paper}.
Both are released under the MIT open-source software license.
This version was generated at git commit
\texttt{\githash\,(\gitdate)}.
This research additionally utilized:
    \texttt{Astropy} (\citealt{Astropy-Collaboration:2013}),
    \texttt{emcee} (\citealt{Foreman-Mackey:2013}),
    \texttt{IPython} (\citealt{Perez:2007}),
    \texttt{matplotlib} (\citealt{Hunter:2007}),
    and \texttt{numpy} (\citealt{Van-der-Walt:2011}).}

\facility{\sdssiii, \apogee}

\bibliographystyle{aasjournal}
\bibliography{thejoker}

\end{document}

%% file: gitstuff.tex
\newcommand{\githash}{3189dc9}\newcommand{\gitdate}{2017-01-30}

%% file: aastexmods.tex
\hypersetup{colorlinks = false}
\makeatletter 
\long\def\frontmatter@title@above{
\vspace*{-\headsep}\vspace*{\headheight}
\noindent\footnotesize
{\noindent\footnotesize\textsc{\@journalinfo}}\par
{\noindent\scriptsize Preprint typeset using \LaTeX\ style AASTeX6
with modifications by DWH and DFM.
}\par\vspace*{-\baselineskip}\vspace*{0.625in}
}%
\long\def\frontmatter@abstractheading{%
\makeaffils
  \vspace*{-\baselineskip}\vspace*{1.5pt}
  \vspace*{0.13189in}
 \begingroup
  \centering
  \abstractname
  \vskip 1mm
  \par
 \endgroup
 \everypar{\rightskip=0.0in\leftskip=\rightskip}\par
}%
\def\frontmatter@keys@format{\vspace*{0.5mm}%
  \settowidth{\keys@width}{\normalsize\@keys@name}%
  \rightskip=0.0in\leftskip=\rightskip\parindent=0pt%
    \hangindent=\keys@width\hangafter=1\normalsize\raggedright}%
\def\twodigits#1{\ifnum#1<10 0\fi\the#1}
\def\mydate{\leavevmode\hbox{\the\year-\twodigits\month-\twodigits\day}}
\makeatother
\renewcommand{\today}{\mydate}

\makeatletter
\let\origsection\section
\renewcommand\section{\@ifstar{\starsection}{\nostarsection}}
\newcommand\nostarsection[1]{\sectionprelude\origsection{#1}}
\newcommand\starsection[1]{\sectionprelude\origsection*{#1}}
\newcommand\sectionprelude{\vspace{1em}}
\let\origsubsection\subsection
\renewcommand\subsection{\@ifstar{\starsubsection}{\nostarsubsection}}
\newcommand\nostarsubsection[1]{\subsectionprelude\origsubsection{#1}}
\newcommand\starsubsection[1]{\subsectionprelude\origsubsection*{#1}}
\newcommand\subsectionprelude{\vspace{1em}}
\makeatother

\sloppypar
